\documentclass[aps,prl,twocolumn,showpacs,groupedaddress]{revtex4-1}

\def\comment#1{\bigskip\hrule\smallskip#1\smallskip\hrule\bigskip}   
\usepackage{amsmath}
\usepackage{graphicx}
\usepackage{dcolumn}
\usepackage{bm}
\usepackage{color}
\usepackage[normalem]{ulem}

\newcommand{\sqw}{{\it{}sqw}}

\begin{document}

\title{Universality in molecular halo clusters}

\author{P. Stipanovi{\'c}$^{1}$, L. Vranje\v{s} Marki{\'c}$^{1,2}$, I.
Be\v{s}li{\'c}$^{1,3}$,
J. Boronat$^3$}
\affiliation{
$^1$ Faculty of Science, University of Split, HR-21000 Split,
Croatia}
 \affiliation{$^2$ Department of Physics and Astronomy, University of Delaware, Newark, USA }
 \affiliation{$^3$ Departament de F\'\i sica i Enginyeria Nuclear, Campus Nord
 B4-B5, Universitat Polit\`ecnica de Catalunya, E-08034 Barcelona, Spain}
 
\def\comment#1{\bigskip\hrule\smallskip#1\smallskip\hrule\bigskip}   
\def\qr{{\bf r}}

\date{\today}

\begin{abstract}
Ground state of weakly bound dimers and trimers with a radius extending
well into the classically forbidden region is explored, with the goal to
test the predicted universality of 
quantum halo states. The focus of the study are molecules consisting of 
T$\downarrow$, D$\downarrow$, $^3$He, $^4$He and alkali atoms, where
interaction between particles is much better known than in the case of
nuclei, which are traditional examples of quantum halos. The study of
realistic systems is supplemented by model calculations in order
to analyze how low-energy properties depend on the
interaction potential. The use of variational and diffusion Monte Carlo
methods enabled very precise calculation of both size and binding
energy of the trimers. In the quantum halo regime, and for
large values of scaled binding energies, all clusters follow almost the same
universal line. As the scaled binding energy decreases, Borromean states
separate from tango trimers.
\end{abstract}

\pacs{}

\maketitle
Universality is important in nearly all areas of physics, since it enables
the establishment of connections between phenomena at different energy and
length scales. It is also a key characteristic of quantum halo states,
usually defined as bound states which extend far into the classically
forbidden regions~\cite{jensen,riisager}.  They were first
recognized~\cite{tanihata1,tanihata2} and traditionally explored in
nuclear physics~\cite{jensen,riisager,frederico12}, but are also known to
exist in molecular physics and have been recently created in ultracold gases 
using Feshbach resonances~\cite{chin10}. Known halo dimers extend over an energy
scale of 16 orders of magnitude. Universality  means that the details of
the potential do not matter, rather all properties of 
dimers can be expressed in terms of the scattering length $a$.  

The concept of universality and quantum halo states was extended to systems with more
particles~\cite{jensen,braaten,frederico}. As in the case of dimers, 
if universality exists the properties of the system are describable by any interparticle
potential where one or a few scattering parameters are the same. It became
clear very soon that the radial extension of the cluster 
is a fundamental quantity which can be used to characterize the states. In order to analyze 
systems across different physics fields dimensionless scaling variables were introduced
and scaled size and scaled energy compared~\cite{riisager00,Jensen03}. The
study of Jensen \textit{et al.} suggests that scaling of trimers 
is approximately universal~\cite{Jensen03}. Universality is expected also in excited Efimov states \cite{Jensen03}. However, such comparisons included
mostly models of nuclear systems which are assumed to be separable into a structureless core and one or more halo particles. Realistic molecular systems are lacking, but, at the same time, interactions between atoms in weakly bound clusters are much better known
than in nuclear systems. Thus, molecular systems can be regarded as a bridge
between nuclear halo states and halo states which appear in ultracold
gases. In this Letter, we show that molecular clusters are the best suited 
systems for testing the universality of scaling between energy and
size of quantum halo clusters.

In 2005  Jensen \textit{et al.}~\cite{jensen} 
predicted a number of molecular systems which could be candidates for quantum halo states.
However, at that time data for both energy and size
of the clusters were available for only $^4$He$_2$, $^4$He$_2$$^3$He and the
excited state of $^4$He$_3$, whose quantum halo character was thus confirmed.  In our previous work on small clusters of
helium and spin-polarized tritium (T$\downarrow$) we also predicted a number of
possible quantum halo clusters~\cite{Stipanovic}. Several studies of the (T$\downarrow$)$_3$  and He-alkali 
dimers and trimers revealed weak binding of some of these systems as well~\cite{blume02,Salci04,LiZhang,Klein,YuanLin,Suno09,LiGou,Baccarelli,Suno13}. 

Experimentally, several molecular quantum halos have been detected so far.
 Among them, using diffraction from the nanoscale grating the $^4$He$_2$ 
 dimer~\cite{He_dimer} and  the $^4$He$_2$$^3$He trimer~\cite{He_clusters}. 
 Recently, a He-Li dimer has been detected as
 well~\cite{Tariq}, offering hope that molecular halo systems with more
 than two particles could be observed and their properties measured.
 
In this work, we study the ground state of selected molecular dimers and
trimers, that are  candidates for quantum halo states  due to their small binding energy, 
with the goal to test the universality of the
predicted scaling laws~\cite{riisager00}. 

\begin{figure}[t]
 \centering
 		\includegraphics[width=0.99\columnwidth]{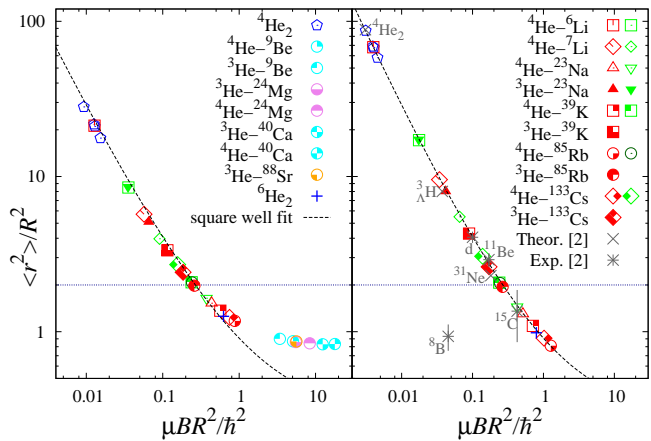}
 	\caption{(Color online) Size \textit{vs} ground-state energy scaling plot for
 	 molecular two-body halos compared with the nuclei data~\cite{riisager}.
	On the left (right) plot $R$ is determined as the outer turning point (the width of the equivalent \sqw{}). The horizontal line 
	is the quantum halo limit and the dashed one the fit through \sqw{} model results. 
	Labels are valid for both sides of the plot. 
	For $^4$He$_2$ the symbols from left to right correspond to interactions from 
	\cite{TTY}, \cite{Aziz}, \cite{SAPT-SM} and \cite{SAPT}. For He-alkali dimers left (right) symbol corresponds to 	interaction from \cite{Cvetko}  (\cite{Klein}).
	The size of the symbols is larger than the errorbar. }
 	\label{fig:dimers}
 \end{figure}

In order to introduce the scaling variables we start the discussion
with the dimers. One has to introduce a length scale 
$R$ with which to compare the size of the dimer, usually quantified through the
root-mean-square radius, $\sqrt{\langle r^2 \rangle}$, with $r$ the distance between the particles.  
In the first model,  $R$ is identified with the outer classical turning
point~\cite{riisager94}. In this case, one can define quantum halo as a
two-body system with a probability to be in a classically
forbidden region higher than 50\%, or as is commonly stated 
$\langle r^2 \rangle/R^2 > 2$. The other
variable is the binding energy, in the scaled form $\mu BR^2/\hbar^2$,
where $B$ equals the absolute value of the ground-state energy and $\mu$ is
the reduced mass of the dimer. We solved numerically the Schr\"odinger
equation for He-He, He-alkali and He-alkaline-earth systems, using several
interaction potentials~\cite{Aziz,SAPT,SAPT-SM,TTY,Klein,Cvetko,Klein00}. The results
for the realistic dimers are presented in Fig. \ref{fig:dimers} as points,
while the line corresponds to the fit through square-well (\sqw{}) model calculations. Our results are in
agreement with the published values of dimer energies from other authors~\cite{Klein,Roudneev2012,LiSong,LiHuang},
while the prediction of $\langle r^2 \rangle$ is 
usually not given. The most
notable example of molecular halo dimer is $^4$He$_2$. Different
models for He-He interactions give binding energies 
from $-1.88$ mK in the 
case of the SAPT potential~\cite{SAPT} to $-1.29$ mK for the TTY
one~\cite{TTY}. Further examples of
halo dimers are the He-alkali systems, the most extended being $^4$He-$^6$Li.  
All studied dimers follow the same curve, even slightly below the
quantum-halo limit, indicated by the horizontal line. As the scaled energy
is even more increased, the  \sqw{} model clearly differs from the
realistic molecular clusters, represented in this energy range by
He-alkaline-earth dimers.  
 
 Since this definition of scaling size $R$ cannot be straightforwardly
extended to systems of more particles, Fedorov \textit{et al.}~\cite{Fedorov94} proposed to 
 define $R$ as the radius of the
equivalent  \sqw{}  potential (e-\sqw), which has the same s-wave scattering length 
and effective range as the original potential. We determined the scattering
length and effective range  for
all He-He and He-alkali systems  under study and, from them, we built 
the e-\sqw{} potentials. Our results are presented
in the right hand side plot of Fig. \ref{fig:dimers}, where the results with \textit{real} and e-\sqw{} potentials are equal.  The scaled size of
quantum halos is somewhat larger in this case. All of the molecular dimers
lie on the line fitted through the  \sqw{} models: e-\sqw{} and modified \sqw{} (m-\sqw) models. The latter have the mass of molecular clusters, but modified depth and width with respect to the e-\sqw. Notably, this holds even below the
quantum halo limit $\langle r^2 \rangle/R^2 = 2$.  The values for the nuclei taken 
from \cite{riisager} in most cases also
follow the universal line. The exception is $^8$B, outside halo region, where centrifugal and Coulomb barrier due to its \textit{core} + p nature presumably reduce its size~\cite{riisager}.

The second definition of the scaling radius was 
extended to trimers~\cite{jensen}. The size of the system is measured by the
root-mean-square hyperradius $\sqrt{\left\langle \rho^2\right\rangle} $, $\rho$ given by
\begin{equation}
m\rho^2=\frac{1}{M}\sum_{i<k}m_im_k(\textbf{r}_i-\textbf{r}_k)^2 \ ,
\label{hyper1}
\end{equation}
where $m$ is an arbitrary mass unit, $m_i$ the particle mass of species $i$, 
and $M$ the total mass of the system. Generalizing the hyperradius (\ref{hyper1}), 
Jensen \textit{et al.}~\cite{jensen} defined the
size scaling parameter $\rho_0$ as 
\begin{equation}
m\rho_0^2=\frac{1}{M}\sum_{i<k}m_im_kR_{ik}^2 \ ,
\label{hyper2}
\end{equation}
where $R_{ik}$ is the two-body scaling length of the $i$-$k$ system,
which is calculated as the width of the e-\sqw{} potential between 
the $i$ and $k$ species. This definition (\ref{hyper2})
enabled the comparison to two-body halos and the analogous definition of
the quantum halo as  $\left\langle \rho^2\right\rangle /\rho_0^2 > 2$.

In order to test the universality of the quantum halos a very accurate
calculation of the energy and size of these 
extremely extended clusters needs to
be done. Although demanding for weakly bound trimers, 
this goal can be
achieved using the diffusion Monte Carlo method (DMC)~\cite{boro} with
pure estimators~\cite{pures}.  The DMC method
solves, within a stochastic approach, the Schr\"odinger equation 
written in imaginary time. For long simulation times, providing that the 
guiding wave function used for importance sampling  
has non-zero overlap with the exact
ground-state wave function, the exact
ground-state energy of a $N$-body bosonic system can be obtained (within some
statistical uncertainty). We used guiding  wave
functions of Jastrow form, constructed as a product of two-body correlation
functions $F_{ij}(r)$, $\psi(\bm{R})=\prod_{i<j=1}^{n}F_{ij}(r_{ij})$.
For realistic potential models we chose 
\begin{equation}
\label{eq:f(r)}
F_{ij}(r)={1\over r}\exp\left[-\left( \alpha_{ij}/r\right)^{\gamma_{ij}}-s_{ij}r\right]  \ ,
\end{equation}
where $r$ is the interparticle distance, and $\alpha_{ij}$, $\gamma_{ij}$ and $s_{ij}$ are
variational parameters. For the  \sqw{} model we used 
\begin{equation}
\label{eq:fsq(r)}
F_{ij}(r)=\left\{
\begin{array}{cl} 
\frac{\sin(k_{ij}r)}{r} &  r\le L_{ij} \\
\exp\left[\frac{k_{ij}(r-L_{ij})}{\tan(k_{ij}L_{ij})}\right]\frac{\sin(k_{ij}L_{ij})}{r} & r>L_{ij} 
\end{array} \right.
\end{equation}
with variational parameters $k_{ij}$ and
$L_{ij}$. Both in Eq. (\ref{eq:f(r)}) and (\ref{eq:fsq(r)}), the parameters were
optimized using the variational Monte Carlo (VMC) method. 

\begin{figure}[t]
 \centering
 		\includegraphics[width=0.99\columnwidth]{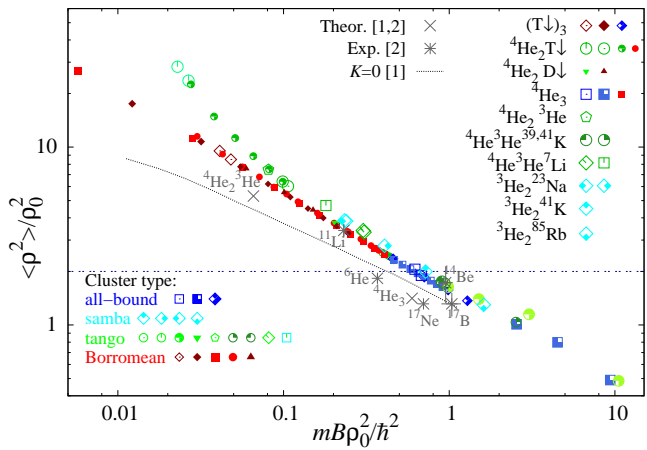}
 	\caption{(Color online) Size \textit{vs} energy scaling plot for ground-state molecular three-body halos. 
	 Empty or 3/4-empty symbols designate 
	 	the realistic, and full or 3/4 full symbols the m-\sqw{} potentials. In the case of $^4$He$_2$T$\downarrow$, 
	left (right) symbol for realistic potential corresponds to He-T interaction from \cite{DWW}, 
	(\cite{MFmod}). For He-alkali trimers, 
	left (right or only one) symbol corresponds to He-alkali interaction from  \cite{Klein}
	(\cite{Cvetko}). Horizontal line represents 
	the quantum halo limit. For comparison, we include the  $K=0$ line
	from \cite{jensen} and  the results from \cite{riisager}. The size of the symbols is larger than the errorbars.}
 	\label{fig:trimers}
 \end{figure}

 Our results for trimers, obtained using interaction potentials \cite{Klein,Aziz,Cvetko,JDW,DWW,MFmod}, are presented in Fig. \ref{fig:trimers} and a sample in more detail in \cite{supplementary}. 
 The empty or 3/4-empty 
 symbols belong to the realistic systems, while the full or 3/4-full ones come from the m-\sqw{} models. Where a molecular system is represented by two equal symbols the left one comes from the
 \textit{real}  two-body potential, and the right one from the e-\sqw.  Here, we see that these two  points lie very
 close, on the same line, and in some cases they are the same within the errorbars. 
 In fact, universality in this context means that clusters 
 can be described by any potential with common
 scattering length and effective range. The
 results for the binding energy of trimers containing only isotopes of He and/or H
 are in good agreement with other published works,~\cite{GN,blume02,Salci04,Suno14,Roudneev2012}. However, previous work,~\cite{jensen} also given in Fig. \ref{fig:trimers} appears to underestimate the sizes of both  $^4$He$_3$ and  $^4$He$_2$$^3$He, placing them below the universal line.
 For the He-alkali trimers, and to the
 best of our knowledge, no results exist with the He-alkali interaction 
 potential by Cvetko  \textit{et al.}~\cite{Cvetko}, while other authors, who used the potential by  Kleinekath\"ofer
 \textit{et al.}~\cite{Klein}, or its older version \cite{KTTY}, modeled
 the He-He interaction with a weaker form than the HFD-B(He)~\cite{Aziz} potential used in 
 the present  work. Thus, we predict somewhat stronger binding for $^3$He$^4$He$^7$Li and
 $^3$He$_2$$^{23}$Na than Yuan and Lin~\cite{YuanLin}. For $^3$He$^4$He$^{39}$K our DMC
 energy is between the lower and upper bounds predicted 
 in  \cite{LiGou}, while for
 $^3$He$_2$$^{85}$Rb we found a bound state only using the potential \cite{Cvetko}, contrary to findings of \cite{LiZhang}.  

\begin{figure*}
\centering
  \includegraphics[width=\textwidth]{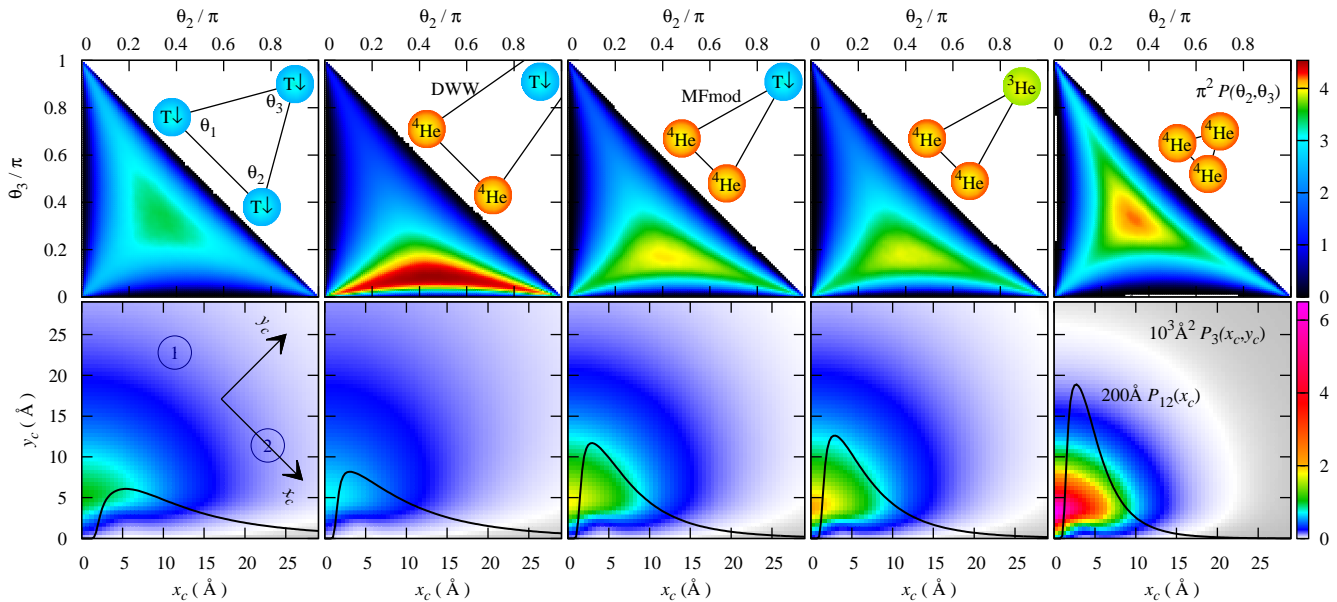}
  \caption{(Color online)Angular probability distribution function
$P(\theta_2,\theta_3)$ in top row and spatial distributions in bottom
row: of half-separations between 1st and 2nd constituent of the trimer
($P_{12}(x_c) \equiv P(r_{12}/2)$; and of positions of the 3rd
constituent in the plane of the trimer where $x_c$ starts from the
center $(\mathbf{r}_2+\mathbf{r}_1)/2$ and points in the direction of
$\mathbf{r}_2-\mathbf{r}_1$.}
  \label{fig:distr}
\end{figure*}

Different types of trimer states, marked with symbols in Fig. \ref{fig:trimers}, are possible: Borromean \cite{jensen},
where no two-body subsystem is bound; tango \cite{tango}, where only one subsystem
is bound; sambas \cite{samba}, with two bound dimers; and all-bound with three bound dimer subsystems.  The only realistic molecular Borromean trimer, (empty diamond) is (T$\downarrow$)$_3$, confirming the predictions of \cite{blume02,Salci04,Beslic08,Suno14}. 
We  obtained other Borromean clusters using m-\sqw{} of $^4$He$_3$, $^4$He$_2$D$\downarrow$, $^4$He$_2$T$\downarrow$ mass, with reduced He-He and strengthened He-H interaction potentials. All of these Borromean states
have $\left\langle \rho^2\right\rangle /\rho_0^2 > 2$ and fall on the same line. Increasing the
interaction strength, the Borromean line passes smoothly into the line of
clusters which have all of their pairs bound. The
only realistic cluster of this type we studied is 
$^4$He$_3$ (empty square), which is exactly
on the border of quantum halo states. Other points represent clusters of $^4$He$_3$ or (T$\downarrow$)$_3$  mass interacting with m-\sqw{} potentials.

 Realistic tango clusters are
$^4$He$_2$$^3$He, $^4$He$_2$T$\downarrow$, $^4$He$^3$He$^7$Li, $^4$He$^3$He$^{39}$K and
$^4$He$^3$He$^{41}$K and all fall approximately on the same line. Not all
tango states are quantum halos, and the criterion $\left\langle \rho^2\right\rangle /\rho_0^2 > 2$
includes $^4$He$_2$$^3$He, $^4$He$_2$T$\downarrow$ and $^4$He$^3$He$^7$Li. Other tango states were again obtained using m-\sqw{} models. As the
binding is reduced, the tango line separates from the Borromean line, 
in accordance with the prediction of Frederico \textit{et
al.}~\cite{frederico}, obtained using renormalized zero-range two-body
interactions. According to that work~\cite{frederico},  
for a given energy the size of the system
increases going from Borromean, through tango, samba and finally all bound
states. Among studied clusters we did not find that samba or
all-bound states separate from the joint Borromean and tango curve.  However, it is not theoretically excluded that this separation would appear for significantly different mass compositions. Samba
clusters $^3$He$_2$$^{23}$Na and $^3$He$_2$$^{41}$K can be considered
quantum halos.
 
Comparing our results with experimental values for nuclei, we find
excellent agreement for $^{11}$Li which is a Borromean state. Other nuclei
fall below the  $\left\langle \rho^2\right\rangle /\rho_0^2 > 2$ limit, however $^{14}$Be and $^{17}$B are within the errorbar of the line formed by molecular clusters. $^{6}$He (two neutrons in p-orbits) and $^{17}$Ne
(\textit{core} + p + p) have reduced sizes with respect to the universal
line. In fact, this is expected because the universal law here obtained is
constrained to s-wave dominated pairwise interactions, without Coulomb
forces.

Let us note that molecular trimer halos appear clearly above the '$K$=0' line in Fig. \ref{fig:trimers}, where $K$ is the hypermoment. This confirms the analysis of \cite{Jensen03}, and the importance of exactly solving the Schr\"odinger equation.

 It is worth noticing that  the choice of the scaling parameter $\rho_0$ is not unique. 
  Jensen
 \textit{et al.}~\cite{jensen03} proposed also a second definition, 
 based on the analysis of the  \sqw{} model, where the 
 mass $m$ is substituted by its square root $\sqrt{m}$.  However, with this definition
 we found that the scaling appears less universal, that is systems with
 different masses are slightly shifted. 

We also studied other structural properties of the clusters, including
their shape and size. In Fig. \ref{fig:distr}, are shown the angular and spatial probability 
distribution functions. On the top plot is also a sketch of the most probable triangle structure with $\left<r_{12}\right>$ divided by
60$\pi^{-1}$\AA{}. 
The angular probability distributions of Borromean and all bound
clusters (at the edges of the plot) have the same symmetry, as expected
because they are constructed from the same  type of particles. However, the less
bound (T$\downarrow$)$_3$  is larger and more spread among different shapes. Angular distributions of the three
tango states, in the middle of the plot, differ from the distributions of the
Borromean or all-bound states. The distributions of $^4$He$_2$$^3$He and
$^4$He$_2$T$\downarrow$  with the MFmod potential \cite{MFmod}, which are close on the universal
plot, appear very similar, which is also the case if one calculates the
weights of different configurations (linear, isoceles, scaline,
equilateral). $^4$He$_2$T$\downarrow$, using DWW \cite{DWW} potential is more weakly bound and larger.
In particular, T is more separated from the $^4$He$_2$ than in the case of the MFmod potential, which can be seen both from  $P(\theta_2,\theta_3)$ and $P(x_c,y_c)$ distributions.

Summarizing, we studied a rather complete set of molecular halo clusters solving the
Schr\"odinger equation in an exact way for both dimers and trimers. In the case of dimers, we
identified the best scaling variables, both in energy and size, which allow for a
universal line on top of which all molecular halo states stand. The analysis of the trimers
is richer due to the different types of halo states one defines according to the
bound or unbound pairs in which a triplet can be decomposed. For the first time, we were
able to establish both the more convenient scaling variables and the universal line which
trimer halo states do follow. The achievement of this universal behavior was possible due
to the accuracy of the interatomic potentials used. Previous attempts of tracing this scaling
law in nuclear systems were not possible due to the approximate validity of the few-body approach and the complexity of nucleon-nucleon potentials.
It is remarkable, and probably unexpected, that the universal law extends even significantly
 below the halo limit  for both dimers and trimers. Finally, we were able to observe, and
 determine when, tango universal line departs from the Borromean one as predicted by
 Frederico \textit{et al.}~\cite{frederico}. 

L.V.M., P. S. and I. B. acknowledge
support from MSES (Croatia) Grant No. 177-1770508-0493 , and L.V.M. partial support from the Fulbright program. J.B. acknowledges partial 
financial support from the DGI (Spain) Grant No.~FIS2011-25275  and Generalitat de Catalunya Grant No.~2009SGR-1003.
The computational resources of the Isabella
cluster at Zagreb University Computing Center (Srce), the HYBRID 
cluster at the University of Split, Faculty of Science and Croatian National
Grid Infrastructure (CRO NGI) were used.

\end{document}